\documentclass[conference]{IEEEtran}
\IEEEoverridecommandlockouts
\usepackage{cite}
\usepackage{amsmath,amssymb,amsfonts}
\usepackage{algorithmic}
\usepackage{graphicx}
\usepackage{textcomp}
\usepackage{xcolor}
\def\BibTeX{{\rm B\kern-.05em{\sc i\kern-.025em b}\kern-.08em
    T\kern-.1667em\lower.7ex\hbox{E}\kern-.125emX}}
\begin{document}

\title{Training Overhead Ratio: A Practical Reliability Metric for Large Language Model Training Systems
}

\author{\textbf{Ning Lu$^{1,2,3,*}$\thanks{$^*$Correspondence to \textit{nluab@cse.ust.hk}.}~\;~Qian Xie$^{1}$~\;~Hao Zhang$^1$~\;~Wenyi Fang$^1$~\;~Yang Zheng$^1$~\;~Zheng Hu$^1$~\;~Jiantao Ma$^4$}\\
    $^1$TTE Lab, Huawei Technologies Co., Ltd.~\;~\\
    $^2$Southern University of Science and Technology~\;~
    $^3$Hong Kong University of Science and Technology~\;~\\
    $^4$Huawei Computing, Huawei Technologies Co., Ltd.
}

\maketitle

\begin{abstract}
Large Language Models (LLMs) are revolutionizing the AI industry with their superior capabilities. Training these models requires large-scale GPU clusters and significant computing time, leading to frequent failures that significantly increase training costs.
Despite its significance,  this field lacks a metric for evaluating reliability. 
In this work, we introduce a novel reliability metric called \emph{Training Overhead Ratio} (TOR) to evaluate the reliability of fault-tolerant LLM training systems.  TOR is defined as the ratio of optimal training time to the observed training time of a system, serving as a practical tool for users to estimate the actual time required to train an LLM on a given system.
Furthermore, our investigation identifies the key factor for enhancing reliability and present TOR equations for various types of failures encountered in practice.


\end{abstract}


\section{Introduction}
Large Language Models (LLMs) have revolutionized the field of artificial intelligence, demonstrating remarkable capabilities across a wide range of tasks, including machine translation, text summarization, and conversational agents~\cite{GPT3, palm}. 
The success of LLMs is primarily attributed to the scaling of model size and training data size, as evidenced by the scaling law~\cite{scaling_law}.
To enhance model capacity, significant efforts have been invested in training increasingly larger models on trillions of tokens, requiring massive GPU clusters over extended periods. 

This unprecedented scale of LLM training presents significant challenges in terms of system reliability.
The month-long duration of training makes failures virtually inevitable.
Moreover, the distributed and parallel nature of the training process exacerbates this issue, as a failure in a single computational node can halt the entire training system. 
For example, the Llama-3 training experienced 466 interruptions during a 54-day snapshot period~\cite{llama3}. 
These frequent failures extends the training time, creating a huge gap between the expected failure-free cost and actual one in practice.

Despite the importance of reliability in LLM training systems, there is no widely accepted metric to effectively measure it. 
Traditional reliability engineering metrics, such as reliability and availability~\cite{availability}, fail to accurately model the LLM training systems.
Reliability, typically modeled by failure rate, does not account for the overhead introduced by reliability improving operation such as checkpointing. 
Availability, calculated as the proportion of failure-free time, is not suitable for LLM training which requires a long time continuous computation. A system might achieve perfect availability by running indefinitely, yet be unreliable due to frequent data corruption.
Another commonly used metric for LLM training system is Model FLOPs Utilization (MFU)~\cite{palm}. 
However, it is designed to measure the system's efficiency, omitting the time waste, such as the interval between checkpoint and failure occurrence.
This gap highlights the need for a more comprehensive and practical metric to measuring the reliability of LLM training systems.

In this paper, we introduced a new metric \textbf{Training Overhead Ratio} (TOR).
It is defined as the ratio of the optimal training time to the observed training time of a system, where the optimal training time refers to the duration necessary to execute a task on an ideal, failure-free system devoid of any operational overhead.
It considers all possible events that extends the training period, e.g., time wasted for roll-back, checkpoint saving overheads.
Using this metric, system users can estimate how long a training time will cost on a given system.
Furthermore,  our investigation show that  performance preservation ratio is crucial for improving the system’s reliability. And we present TOR equations for various types of failures encountered in practice.

In summary, our contributions are: 
1) We propose the first reliability metric for LLM training systems and provide a detailed equation for its calculation.
2) We identify key factors that enhance system reliability through investigation.

\section{Training Overhead Ratio}

\subsection{Definition}
We introduce a novel metric, termed the \textbf{Training Overhead Ratio} (TOR), designed to evaluate the reliability of fault-tolerant LLM training systems. 
TOR is defined as the ratio of the optimal training time to the actual observed training time, offering a quantitative assessment of a system's performance relative to its failure-free efficiency.
Optimal training time is characterized as the duration necessary to execute a task on an ideal, failure-free system devoid of any operational overhead.
In simpler terms, it represents the uninterrupted execution time of a task.
It is mathematically represented as follows:
\begin{equation}
\label{eq:metric}
\text { Training Overhead Ratio }=\frac{\text { Optimal Training Time }}{\text { Observed Training Time }}.
\end{equation}
TOR ranges from 0 to 1, with higher values indicating a greater reliability of the system. 
It gives empirical comparisons among different training systems, enabling users to estimate the actual time required to train an LLM on a given system.


\subsection{Performance Preservation Ratio}


In this section, we introduce the concept  of performance preservation ratio and demonstrates its crucial role for improving the system's TOR.

In any training task, the total workload, e.g., the number of FLOPs, is fixed. 
The equality for the computational effort between the optimal training and the failure-aware one is captured by the following equation:
\begin{equation}
\label{eq:total_work}
 W_{\text{opt}} \cdot T_{\text{opt}} = \int_{0}^{T_{\text{obs}}} w_{\text{obs}}(t) \, dt,
\end{equation}
where $w_{\text{obs}}(t)$ represents the observed work rate at time $t$. For optimal training, we simplify the optimal work rate as a constant $W_{\text{opt}}$.
Note that the optimal work rate does not imply optimal hardware utilization, e.g., 100\% MFU. Instead, it represents the normal working rate without failures.  
$T_{\text{opt}}$ and $T_{\text{obs}}$ denote the optimal and observed training time, which is required for TOR calculation. 
By dividing $W_{\text{opt}}$ on both sides, Eq.~\ref{eq:total_work} simplifies to:
\begin{equation}
\label{eq:t_relation}
 T_{\text{opt}} = \int_{0}^{T_{\text{obs}}}  \mathbf{r}(t) \, dt,
\end{equation}
where $\mathbf{r}(t) = w_{\text{obs}}(t) / W_{\text{opt}} \in [0, 1]$ is the {performance preservation ratio}, representing the proportion of the optimal work rate achieved in practice at time $t$.

Eq.~\ref{eq:t_relation} provides a tractable method for estimating the optimal training time.
Furthermore, it suggests that enhancing the TOR requires either maximizing $\mathbf{r}(t)$ or minimizing the duration of low $\mathbf{r}(t)$ values during training.

\subsection{Failure Modeling}

To further analyze the impact on the system reliability, or the performance preservation ratio, we develop a simple failure-recovery model based on industrial experience.

Modern LLM training systems encounter various types of failures during the training process~\cite{llama3}. 
We categorize these failures into two types based on their recovery patterns: fail-stop and fail-slow failures.
Fail-stop failures instantly stop the training process and requires roll-back operation for recovery. 
In contrast, fail-slow failures allow the training process to continue but at a lower-than-expected performance level.

\begin{figure}[tbh]
  \centering
  \includegraphics[width=0.5\textwidth]{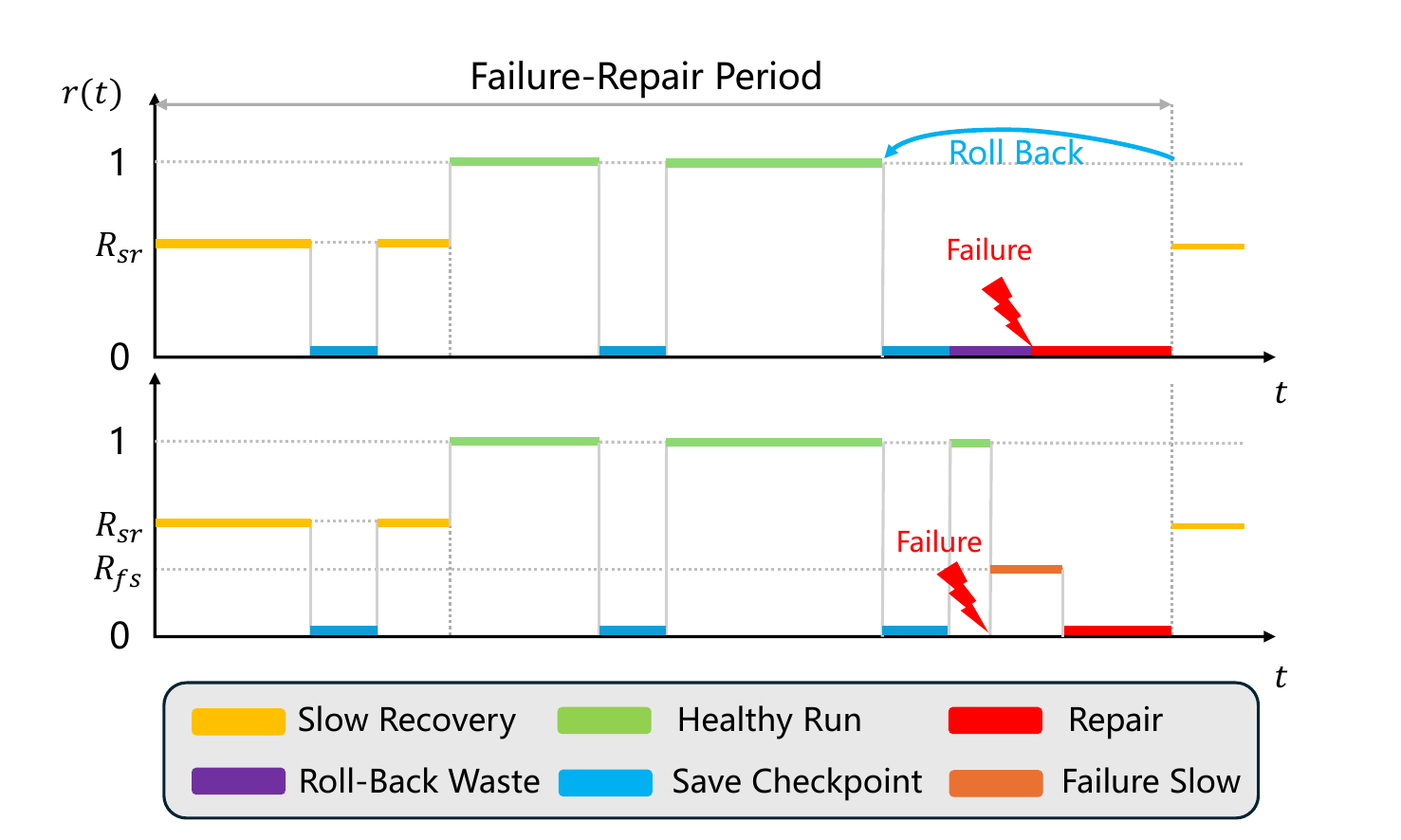} 
  \caption{The changes of performance preservation ratio $\mathbf{r}(t)$ in different status in one failure-repair period, for fail-stop (upper) and fail-slow (lower) failures.}
  \label{fig:failure_model}
\end{figure}

We focus our analysis on a single failure-recovery period, which can be viewed as a repeating unit throughout the training process. Both failure types share the following stages:
\begin{itemize}

\item Slow Recovery: Post-failure recovery where the system operates at a reduced functionality ($\mathbf{r}(t) < 1$) for duration $T_{sr}$, with $\mathbf{r}(t)$ approximated as constant $R_{rs}$ for simplicity. 

\item Healthy Run: Normal system runs without failures, where $\mathbf{r}(t) = 1$ for duration $T_{h}$.

\item Checkpoint Saving: The system halts training for $T_{ckpt}$ to save checkpoints, occurring $N_{ckpt}$ times per period. 

\item Repair: Repair: Training stops for duration $T_r$ to address the failure, where $\mathbf{r}(t) = 0$.

\end{itemize}

For fail-stop failures, the system will roll back to the latest checkpoint, resulting in wasted time $T_{rb}$ between the checkpoint and failure occurrence.  
Although the system works for this period, the work does not contribute to the final model, thus $\mathbf{r}(t) = 0$. 
Figure~\ref{fig:failure_model} illustrates the stages and corresponding $\mathbf{r}(t)$ changes in one failure-repair period. 

Based on this modeling, we can derive the TOR equation for fail-stop failure:
\begin{align}
 \text{TOR}_{\text{fail-stop}} &= \frac{T_{\text{opt}}}{T_{\text{obs}}} 
 = \frac{\int_{0}^{T_{\text{obs}}}  \mathbf{r}(t) \, dt}{T_{\text{obs}}} \notag \\
 &= \frac{T_{sr} R_{rs} + T_h}{T_{sr} + T_h + N_{ckpt}T_{ckpt} + T_{rb} + T_{r}} \\
 &= \frac{\text{MTBF} - T_{sr}(1-R_{sr}) - T_{rb} - N_{ckpt}T_{ckpt}}{\text{MTBF} + T_{r}}. \notag
\end{align}
Notice that the Mean Time Between Failures (MTBF) is calculated by the formula  $\text{MTBF} = T_{sr} + T_h + N_{ckpt}T_{ckpt} + T_{rb}$, which depends on the intrinsic characteristics of a system, such as hardware components and architecture. 

Similarly, the TOR equation for fail-slow failures is 
\begin{align}
 \text{TOR}&_{\text{fail-slow}} 
 = \frac{T_{sr} R_{rs} + T_h + T_{fs} R_{fs}}{T_{sr} + T_h  + N_{ckpt}T_{ckpt} + T_{fs} + T_{r}} \notag\\
 &= \frac{\text{MTBF}  - T_{sr}(1-R_{sr}) - N_{ckpt}T_{ckpt} + T_{fs} R_{fs}}{\text{MTBF} + T_{fs} + T_{r}},  
\end{align}
where $\text{MTBF} = T_{sr} + T_h + N_{ckpt}T_{ckpt}$.

For a system with multiple types of failures, the TOR can firstly be calculated separately for each one. Then these values can then be weighted according to their occurrence rates and summed to obtain the final TOR.

\section{Conclusion}
In this paper, we propose the first reliability metric termed Training Overhead Ratio for LLM training system. 
This is the ratio of the optimal training time to the actual observed training time of a system.
Furthermore, our investigation identifies the key factor for enhancing reliability and present TOR equations for various types of failures encountered in practice.

\bibliographystyle{IEEEtran}
\bibliography{ref}

\end{document}